# A Differential Geometric Context in which Quantization is a Necessary Condition for Gauge Invariance

J. Towe

Department of Physics, The Antelope Valley College
Lancaster, CA, USA
jtowe@avc.edu

It has recently been argued that quantization can be established within classical theory as a consequence of lost information. In this view, classical mechanics is regarded as a union of quantum mechanics and what are called 'hidden variables'. Hidden variable theories were first considered some years ago and abandoned, because they did not produce new physics. They have now been revived however, because they appear to provide a logically sound means of relating classical and quantum theories. It is argued here that the Heisenberg uncertainty relations constitute a necessary condition for gauge invariance in the 5-dimensional Kaluza-Klein theory, where the fifth dimension of a 5-spacetime is hidden as a 5-dimensional theory of general relativity is projected onto 4-spacetime.

## 1. Introduction

Quantum mechanics has superceded all classical theories other than the general theory of relativity (GTR), and indeed, it has traditionally been expected that GTR will ultimately be regarded as an approximation of a quantum theory of gravity. Einstein rejected this view however, and argued that quantum mechanics would ultimately give way to classical physics. It was traditionally a 'hard call', because both GTR and quantum mechanics are incomplete. While GTR cannot account for stellar collapse beyond the singularities that characterize the theory, quantum mechanics cannot account for individual events. But classical theory contains more information than quantum theory, and in this context, it was argued that a classical theory can become a quantum theory if the former, for some reason, loses information. In this view, classical theory is regarded as a union of quantum mechanics and certain 'hidden variables'. The approach involving hidden variables was considered by Einstein, but ultimately discarded by him, and by the physics community, because hidden variable theories produced no new physical results. Because the hidden variable approach appears to be logically sound however, it has been revived. One such theory was proposed by this author. It was demonstrated that the Bohr radii constitute a necessary condition for gauge invariance in the 5-dimensional Kaluza-Klein theory. Specifically, it was demonstrated that if GTR on 5-spacetime is projected onto 4-spacetime, then the Bohr radii constitute a necessary condition for U(1) invariance [J. Towe. 1988]. A second such theory was proposed by B. Muller, who demonstrated that a classical system in five dimensions becomes a quantum system if observed in four dimensions [B. Muller, 2001]. A complementary result is due to R. Bousso, who derived the Heisenberg relation from the holographic limit [R. Bousso, 2004]. Complimenting these results, G. 't Hooft contributed an argument that classical theories may lose information due to dissipative forces [G. 't Hooft, 1999].

 A theory that was similar to the above described efforts was proposed in the context of Hermann Weyl's theory of gravitation and classical electrodynamics [H. Weyl, 1918]. The latter theory postulated a generalization of Riemannian geometry in which vectors can undergo increments both of orientation and scale when parallel transported around a closed curve. The degrees of freedom that result from this hypothesis are precisely sufficient to produce Einstein's gravitation in the astrophysical domain of 4-spacetime and classical electrodynamics in the macroscopic domain. The discovery that the Weyl theory also requires quantization conditions occurred when Einstein critiqued Weyl's theory, observing that the orbital motions of electrons would, in this theory, associate with increments of proper time scale that would contradict observation [H. Weyl, 1922]. This claim motivated F. London to demonstrate that the Klein-Gordon equation can be derived, in the context of the Weyl theory, as a necessary condition for the preservation of proper time under Weyl scale transformations [F. London, 1927]. Although these results are interesting, the theories of Weyl and London were not extensively studied, because they produced no new physical results.

 The theory that is introduced here parallels the theory of London, and constitutes a generalization of the theory which was introduced by this author in 1988. The fifth spacetime dimension is again regarded as a hidden variable and discarded; i.e. general relativity in five dimensions is again projected onto 4-spacetime. The relevance of the proposed generalization of the 1988 theory is that it derives *general* quantization (the Heisenberg relations) as a necessary condition for gauge invariance. The proposed theory also parallels the theory of Muller in that it produces a quantum mechanical version of a classical system by hiding a fifth spacetime dimension, and complements the theory of Bousso in that it establishes the very general Heisenberg relation. The proposed theory should also be compared with the thinking of van de Bruck, who speculated that the dissipative forces, discussed by 't Hooft, may be related to gravity [C. van de Bruck 2000]. But the approach proposed here is strictly in the tradition of Einstein, because it is exclusively differential geometric in character, and derives general quantization conditions from the 5-dimensional theory of general relativity. Finally, it should be observed that the theory proposed here appears to be complemented by recent discussions regarding an ultimate superunification. Specifically, the proposed theory models, from a strictly differential geometric perspective, the (quantum electrodynamical) stationary states of Dirac and Klein-Gordon particles; e.g. of electrons. Based upon recent results, it appears that supergravity in the context of superstring compactification, can be regarded as modeling (supergravitational) stationary states of the baryon [J. Towe, 2005].

## 2. Gauge Invariance in the 5-dimensional Kaluza-Klein Theory

Let us recall the Christoffel connections (of the second kind) that underlie general relativity mechanics on 4-spacetiime:

$$\begin{Bmatrix} \mu \\ \nu \ \ \rho \end{Bmatrix} = \frac{g^{\mu\alpha}}{2} \left( \frac{\partial g_{\alpha\nu}}{\partial x^{\rho}} + \frac{\partial g_{\rho\alpha}}{\partial x^{\nu}} - \frac{\partial g_{\nu\rho}}{\partial x^{\alpha}} \right): \qquad (1)$$

μ, ν, ρ = 1, 2, 3, 4 [A. Einstein, 1915]. An analogous set of connection coefficients characterize the general theory of relativity on 5-spacetime that was considered by Theodor Kaluza and Oscar Klein

$$\begin{Bmatrix} M \\ N \ \ R \end{Bmatrix} = \frac{g^{MA}}{2} \left( \frac{\partial g_{AN}}{\partial x^{R}} + \frac{\partial g_{RA}}{\partial x^{N}} - \frac{\partial g_{NR}}{\partial x^{A}} \right): \qquad (2)$$

M, N, R = 1, 2, 3, 4, 5 [T. Kaluza, 1921; O. Klein, 1926; O. Klein, 1938]:
. The coefficients

$$\begin{Bmatrix} \mu \\ \nu \ \ 5 \end{Bmatrix} = \frac{g^{\mu\alpha}}{2} \left( \frac{\partial g_{\alpha\nu}}{\partial x^{5}} + \frac{\partial g_{5\alpha}}{\partial x^{\nu}} - \frac{\partial g_{\nu 5}}{\partial x^{\alpha}} \right): \qquad (3)$$

μ, ν = 1, 2, 3, 4 clearly describe the connection coefficients that are, in addition to those described by (1), perceived on 4-spacetime in the context of the Kaluza-Klein theory. Moreover, the 'Kaluza-Klein' condition

$$\frac{\partial g_{\alpha\nu}}{\partial x^5} = 0 \tag{4}$$

reduces the coefficients (3) to

$$F^{\mu}{}_{\nu} = \frac{g^{\mu\alpha}}{2}\left(\frac{\partial g_{5\alpha}}{\partial x^{\nu}} - \frac{\partial g_{\nu 5}}{\partial x^{\alpha}}\right), \tag{5}$$

which Kaluza and Klein interpreted as constituting the electromagnetic field tensor on 4-spacetime. Let us now consider the energy of this electromagnetic field in the presence of 4-current. If the Lagrangian, $\mathcal{L}$, is described in terms of $F^{\mu}{}_{\nu}$, if $\mathcal{L}$ is invariant under the group U(1) of 1-parameter phase transformations, and if the variational principle

$$\delta \int \mathcal{L}\, d^4x = 0,$$

is restricted to a variation of action with respect to the metrical coefficients $g_{\mu 5}$, then the Euler-Lagrange equations are the Maxwell equations (that describe classical electromagnetic interactions in 4-spacetime). If these are integrated, the result is a classical electromagnetic wave in 4-spacetime. But this electromagnetic wave is distinguished from the ordinary electromagnetic wave. This wave, due to its origin in the Kaluza-Klein theory, is characterized by special invariance properties. Specifically, this wave is invariant under phase transformations that constitute a group, which can be parameterized in terms of either electrical charge or 4-momentum.

The group U(1) consists (to first order in the parameter $u = u_{\mu}dx^{\mu}$) of the infinitesimal phase transformations

$$\exp\sum_{\mu=1}^{4} u_{\mu}dx^{\mu} I = I + \sum_{\mu=1}^{4} u_{\mu}dx^{\mu}. \tag{6}$$

U(1) is usually parameterized in terms of electrical charge, so that invariance of an interaction under U(1) is, by Noether's theorem, equivalent to conservation by that interaction of electrical charge. In the Kaluza-Klein theory however, this parameterization in terms of electrical charge is equivalent to parameterization in terms of 4-momentum. To demonstrate this, let us consider the tensorial equations describing the acceleration of an electrical charge in an electromagnetic field:

$$m\frac{d^2x^{\mu}}{d\tau^2} = \begin{Bmatrix} \mu \\ \nu\ 5 \end{Bmatrix} e\left(\frac{dx^{\nu}}{d\tau}\right) = eF^{\mu}{}_{\nu}\frac{dx^{\nu}}{d\tau} : \mu = 1, 2, 3, 4, \tag{7A}$$

where e is the electrical charge on the test particle, where m is the mass of the test particle, where $m\frac{d^2x^\mu}{d\tau^2} = eF^\mu{}_\nu \frac{dx^\nu}{d\tau}: \mu = 1, 2, 3, 4$ describes the Lorentz force of an electromagnetic field on the test particle and where we have adopted the Einstein summation convention (summing over repeated Greek indices from 1 through 4, and over repeated Latin indices from 1 through 3). Thus, comparing equation 7A with the equations of motion of a test particle in a gravitational field:

$$m\frac{d^2x^\mu}{d\tau^2} = \begin{Bmatrix} \mu \\ \nu\ \rho \end{Bmatrix} m(\frac{dx^\nu}{d\tau})\frac{dx^\rho}{d\tau} = \begin{Bmatrix} \mu \\ \nu\ \rho \end{Bmatrix} P^\nu \frac{dx^\rho}{d\tau}: \mu = 1, 2, 3, 4, \quad (7B)$$

one observes that electrical charge, e, constitutes the 5[th] component of 5-momentum that is, in the context of the Kaluza-Klein theory, represented by the charge of a test particle in an electromagnetic field. In the Kaluza-Klein context then, 5-action is given by

$$P_5 dx^5 - P_\mu dx^\mu = e dx^5 - P_\mu dx^\mu.$$

But, just as 4-action $Edt - P_j dx^j$ reduces to 3-action $Edt = P_j dx^j$ on 3-spacetime, 5-action $e dx^5 - p_\mu dx^\mu$ reduces to 4-action $e dx^5 = p_\mu dx^\mu$ on 4-spacetime. In the 4-spacetime projection of 5-dimensional GTR, which is required by the Kaluza-Klein theory then, parameterization of the group U(1) of phase transformations in terms of electrical charge $e dx^5$ is equivalent to parameterization in terms of 4-momentum $P_\mu dx^\mu$.

In this context, consider a classical wave propagating in 4-spacetime; and consider the transformation of this wave under the group U(1) of phase transformations:

$$\hat{\Psi}^\alpha = \Psi^\alpha \exp i \frac{P_\mu dx^\mu}{k}, \quad (7)$$

where k is the smallest positive value with dimensions of action. The invariance of this wave state under the relevant group U(1) clearly implies that

$$\hat{\Psi}^\alpha = \Psi^\alpha \exp i \frac{P_\mu dx^\mu}{k} = \Psi^\alpha \exp i(2\pi n) = \Psi^\alpha \cos(2\pi n) = \Psi^\alpha: \nu = 0, 1, 2, 3, ... \quad (8)$$

where $P_4 = -E/ic$, $dx^4 = icdt$; or that

$$\frac{P_\mu dx^\mu}{k} = 2\pi n: n = 0, 1, 2, 3, \ldots \quad (9)$$

Since division by zero is mathematically precluded, and since negative action is not contemplated, it must be concluded that k>0 (as stipulated above), so that

$$P_\mu dx^\mu \geq 2\pi k. \quad (10)$$

Moreover, if axes are aligned so that motion is parallel to the μ-axis, one can write

$$P_\mu dx^\mu \geq 2\pi k, \qquad (11)$$

where summation is not implied. Thus, since every axis can be aligned with motion, one can write

$$P_\mu dx^\mu \geq 2\pi k: \mu = 1, 2, 3, 4. \qquad (12)$$

Finally, since k has the dimensions of action, one can assign to k a constant value:

$$k = \frac{\hbar}{4\pi}, \qquad (13)$$

so that (11) reduces to

$$P_\mu dx^\mu \geq \frac{\hbar}{2}: \mu = 1, 2, 3, 4; \qquad (14)$$

which are the Heisenberg relations on 4-spacetime [L. Schiff, 1968].

## Conclusion

It should be re-emphasized that the above result is possible because the electromagnetic signal under consideration is obtained from the Kaluza-Klein theory, where parameterization of the group U(1) of phase transformations in terms of electrical charge $e dx^5$ is equivalent to parameterization of that group in terms of 4-momentum $P_\mu dx^\mu$ (see equations 7A and 7B).

The notion that quantum mechanics can emerge from classical theory; e.g. from general relativity is often described as infeasible due to the counterintuitive nature of quantum theory. Even though quantum physics contains less information than classical physics, it is argued that the many strange and counterintuitive concepts that are involved in the former can hardly be accounted for in terms of classical physics. It should be recognized however, that one is not obligated to account, one by one, for the details of quantum theory. All of quantum mechanics (up to relativistic considerations) can be derived from the Heisenberg uncertainty principle, or from the logically equivalent Schrodinger formulation. The essential element is simply the non-zero value of the quantum of action.